\documentclass[twocolumn]{aastex62}
\usepackage{xcolor}
\usepackage{bm}
\usepackage[capitalize]{cleveref}

\graphicspath{{./}{plots/}}
\DeclareGraphicsExtensions{.pdf,.eps,.png}

\newcommand{\sect}[1]{Section \ref{sec:#1}}

\definecolor{q}{HTML}{228B22}
\definecolor{wc}{HTML}{FF8C00}
\definecolor{dnc}{HTML}{FF00FF}
\definecolor{todo}{HTML}{e13748}
\definecolor{ben}{HTML}{e13748}

\newcommand{\kms}{\ensuremath{\mathrm{km \, s^{-1}}}}

\newcommand{\sw}{\ensuremath{\mathrm{sw}}}

\newcommand{\vsw}{\ensuremath{v_\sw}}
\newcommand{\vi}{\ensuremath{v_\mathrm{i}}}
\newcommand{\vv}{\ensuremath{v_0}}
\newcommand{\Wind}{\textit{Wind}}
\newcommand{\SWICS}{\textit{SWICS}}

\newcommand{\ssn}{\ensuremath{\mathrm{SSN}}}
\newcommand{\ahe}{\ensuremath{\mathrm{A}_\mathrm{He}}}

\newcommand{\days}{\ensuremath{\mathrm{days}}}
\newcommand{\ndays}{\ensuremath{N_\mathrm{days}}}
\newcommand{\xcorr}{\ensuremath{\rho(\ahe,\ssn)}}
\newcommand{\chisqnu}{\ensuremath{\chi^2_\nu}}
\newcommand{\rsq}{\ensuremath{R^2}}

\newcommand{\Lalpha}{\ensuremath{\mathrm{L}\alpha}}

\newcommand{\sxr}{\ensuremath{\mathrm{SXR}}}

\newcommand{\tauS}{\ensuremath{\tau_\mathrm{slow}}}
\newcommand{\tauF}{\ensuremath{\tau_\mathrm{fast}}}
\newcommand{\singlepanelwidth}{\linewidth}

\newcommand{\plotAbundance}{
\begin{figure*}
\includegraphics[width=\linewidth]{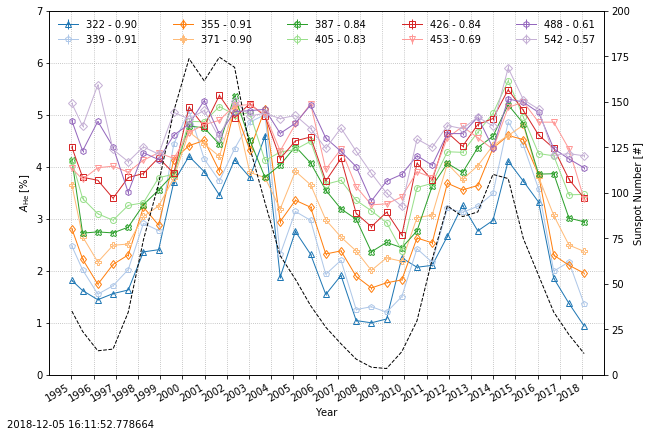}
\caption{Helium abundance (\ahe) as a function of time and solar wind speed.
Solar wind speed (\vsw) is divided into ten quantiles. 
Thirteen month smoothed SIDC Sunspot Number (SSN, dashed black) is plotted on the secondary y-axis.
The legend indicates the middle of a given \vsw\ quantile and the Spearman rank correlation coefficient between \ahe\ and \ssn\ for that quantile.
In effect, this figure updates Fig. (1) of \citet{Kasper2007,Kasper2012}.
The present drop in \ahe\ reflects the onset of solar minimum 25.
\label{fig:abundance_cycle}}
\end{figure*}
}

\newcommand{\plotVswPeakDelay}{
\begin{figure}
\gridline{
    \fig{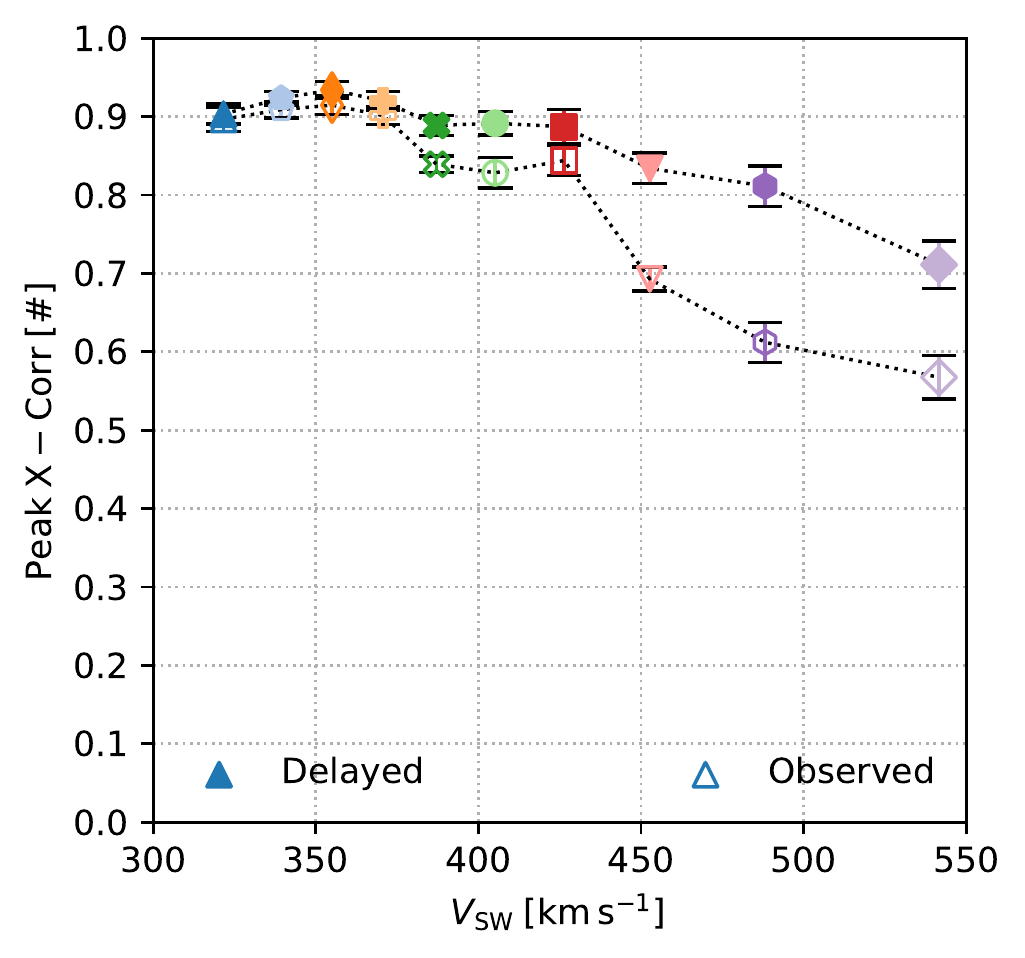}{0.5\textwidth}{(a)}
    }
    \vspace{-2ex}
\gridline{
    \fig{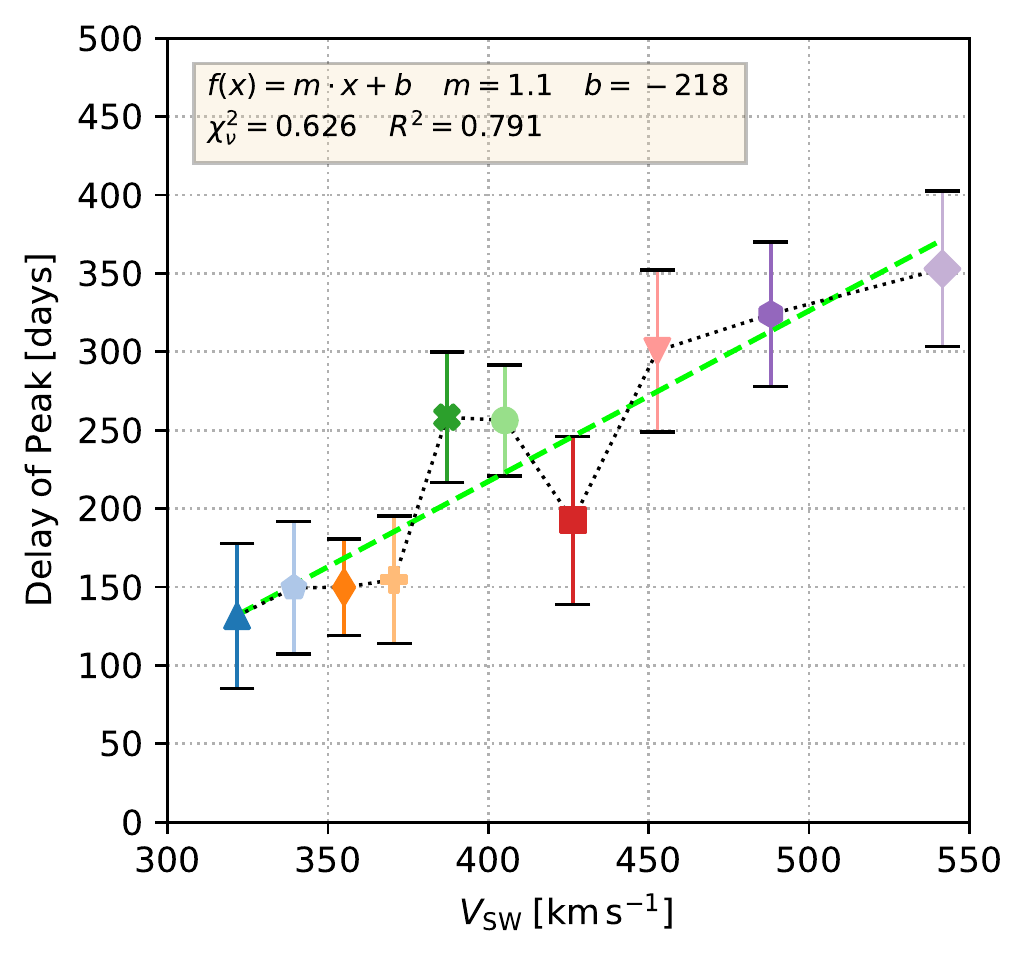}{0.5\textwidth}{(b)}
    }
\caption{
Plots characterizing the cross correlation coefficient as a function of solar wind speed (\vsw) for the observed (empty markers) and delayed (filled markers) \ssn\ using 250 day averages.
The marker color and shape match the style of \cref{fig:abundance_cycle}.
Dotted lines connect the markers to aid the eye.
Panels are: (a) Spearman rank cross-correlation coefficient and (b) Delay \added{($\tau$)} of Peak Spearman rank cross correlation \deleted{($\tau$)} as a function of \vsw.
In (b), the dashed green line indicates a robust fit and the panel's insert provides the functional form, fit parameters, and quality metrics.
A positive delay indicates that changes in \ssn\ precede changes in \ahe.
\label{fig:vsw_xcorr_delay_ebar}}
\end{figure}
}

\newcommand{\MultiPanelWidth}{0.2\textwidth}
\renewcommand{\MultiPanelWidth}{0.5\textwidth}
\newcommand{\plotHysteresis}{
\begin{figure}
\gridline{
          \fig{hysteresis/vsw-355/obs}{\MultiPanelWidth}{(a)}
          }
          \vspace{-2ex}
\gridline{
          \fig{hysteresis/vsw-355/delay}{\MultiPanelWidth}{(b)}
          }
\caption{
Helium abundance (\ahe) as a function of (a) observed and (b) delayed \ssn\ in one example $\vsw$ quantile.
A line connects the points to aid the eye.
Line and marker color correspond to the number of days since mission start.
\added{Marker shape matches the quantile in previous figures.}
This \vsw\ quantile covers the range \replaced{$415 \; \kms < \vsw \leq 438 \; \kms$}{$347 \; \kms < \vsw \leq 363 \; \kms$}.
A green, dashed line presents a robust fit to each trend.
The insert at top of each panel gives the function fit, fit parameters, and quality metrics.
Delaying \ssn\ by the phase offset appropriate to this \vsw\ quantile reduces the impact of the hysteresis effect, as the increase in delayed \rsq\ indicates.
\added{That \chisqnu\ is closer to unity in (b) indicates that a linear model better describes \ahe\ as a function of delayed \ssn.}
\label{fig:hysteresis}}
\end{figure}
}

\newcommand{\plotHysteresisSummary}{
\begin{figure}[h]
\includegraphics[width=\singlepanelwidth]{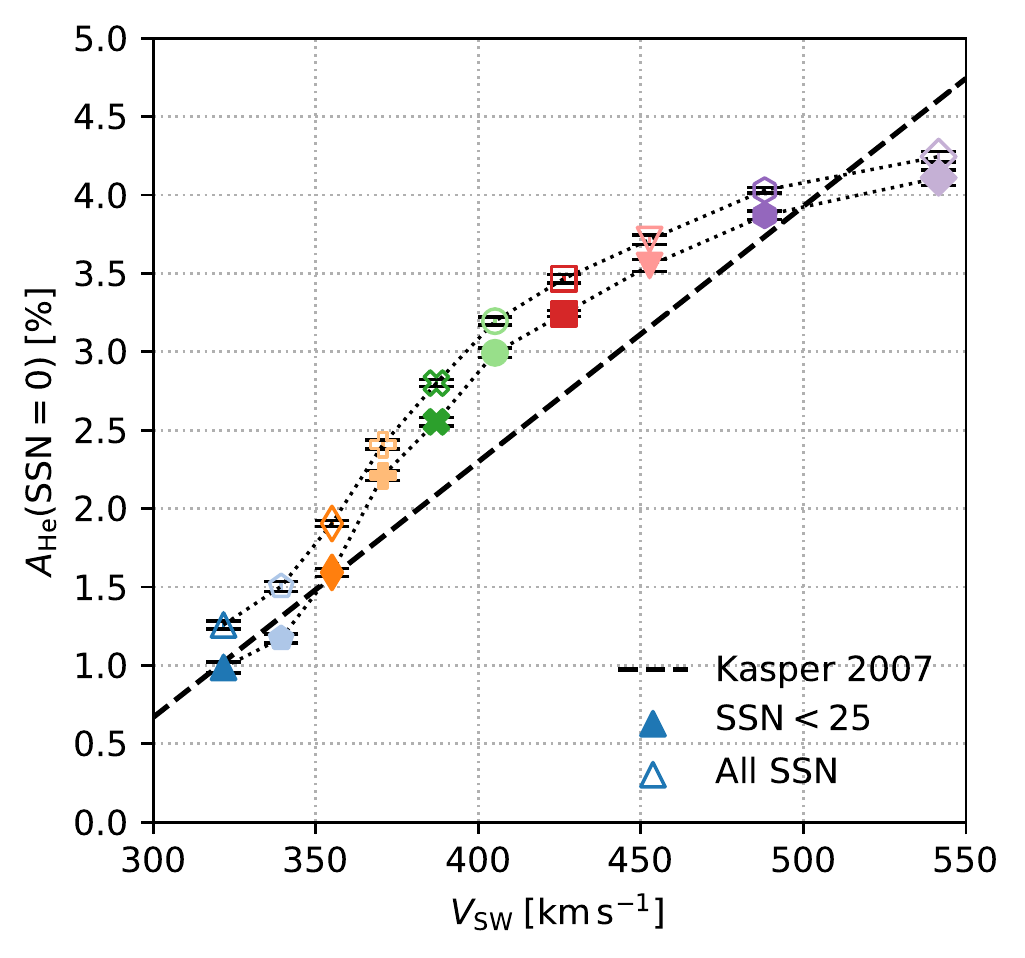}
\caption{A summary of the zero solar activity helium abundance, $\ahe (\ssn=0)$, as a function of \vsw\ for all robust fits in the fashion of \cref{fig:hysteresis}.
Error bars indicate the standard deviation of each quantity over the range in averaging windows $225 \leq \ndays \leq 275$ , each centered on the $\ndays = 250$ value.
Unfilled markers show all \ssn.
Filled markers show identical calculations with $\ssn < 25$.
The black dashed curve is the relationship between \ahe\ and \ssn\ derived by \citet{Kasper2007}.
That repeating our calculation with a reduced range in \ssn\ shows better agreement with the results of \citet{Kasper2007} indicates that our results, covering the full range of solar activity in cycles 23 and 24, are consistent with their results from the two year interval surrounding minimum 23.
\label{fig:hysteresis-summary}}
\end{figure}
}

\watermark{Published}

\begin{document}
\title{Helium Variation Across Two Solar Cycles \added{Reveals A Speed-Dependent Phase Lag}}

\shorttitle{Helium}
\shortauthors{Alterman et al.}

\correspondingauthor{B. L. Alterman}
\email{balterma@umich.edu}

\author[0000-0001-6673-3432]{B.\ L.\ Alterman}
\affiliation{University of Michigan \\
Department of Applied Physics\\
450 Church St.\\
Ann Arbor, MI 48109, USA}
\affiliation{University of Michigan \\
Department of Climate \& Space Sciences \& Engineering \\
2455 Hayward St. \\
Ann Arbor, MI 48109-2143, USA}

\author[0000-0002-7077-930X]{Justin C.\ Kasper}
\affiliation{University of Michigan \\
Department of Climate \& Space Sciences \& Engineering \\
2455 Hayward St. \\
Ann Arbor, MI 48109-2143, USA}
\affiliation{Smithsonian Astrophysical Observatory \\
Observatory Building E \\
60 Garden St.\\
Cambridge, MA 02138, USA}




\begin{abstract}

We study the relationship between solar wind helium to hydrogen abundance ratio (\ahe),  \added{solar wind speed (\vsw),} and sunspot number (\ssn) over solar cycles 23 and 24.
This is the first full 22-year Hale cycle measured with the \Wind\ spacecraft \replaced{showing one complete rotation or cycle of the solar dipole}{covering a full cycle of the solar dynamo with two polarity reversals}.
While previous studies have established a strong correlation between \ahe\ and \ssn, \replaced{here we show a consistent speed-dependent offset between the two quantities}{we show that the phase delay between \ahe\ and \ssn\ is a monotonic increasing function of \vsw}.
Correcting for this lag, \ahe\ returns to the same \added{value at a given} \ssn\ over all rising and falling phases and across solar wind speeds.
\added{We infer that this speed-dependent lag is a consequence of the mechanism that depletes slow wind \ahe\ from its fast wind value during solar wind formation.}

\end{abstract}


\section{Introduction} \label{sec:intro}

\plotAbundance

Fully ionized hydrogen or protons ($p$) and fully ionized helium or alpha particles ($\mathrm{He}$ or $\alpha$) are the two most abundant solar wind ion species.
The former comprises $\sim 95\%$ of the solar wind ions and the later $\sim 4\%$, both by number density. 
Heavier, minor ions constitute the remaining.
The alpha particle abundance ($\ahe = 100 \times n_\alpha/n_p$) strongly correlates with solar activity, as indicated by the sunspot number (\ssn) \citep{Aellig2001,Kasper2007,Kasper2012}.
The cross correlation and slope between \ahe\ and \ssn\ varies with solar wind speed (\vsw); is strongest in slow wind; markedly falls off above $\vsw = 426 \;\kms$, where \ahe\ takes on a stable value between $4\%$ and $5\%$; and vanishes in the solar wind for speeds below $\vv = 259 \; \kms$ \citep{Kasper2007,Kasper2012}.
This helium vanishing speed is within $1 \; \sigma$ of the minimum observed solar wind speed \citep{Kasper2007}, indicating that helium may be essential to solar wind formation in the corona.

In addition to \ssn, many other indicators of solar activity also follow a similar $\sim 11$ year cycle \replaced{. [and references therein]{Ramesh2014} Many of these indices}{\citep{Ramesh2014}  that} demonstrate a distinct phase-offset with \ssn, which has been referred to as a hysteresis-like effect.
\added{These offsets range from 30 days \citep{Bachmann1994} to 450 days \citep{Temmer2003}.}
\citet{Goelzer2013} have shown a similar phase lag in the interplanetary magnetic field's response to changes in \ssn.

Using observations from the \Wind\ Faraday cups (FC), we extend the study of \ahe\ variation with \ssn\ and \vsw\ by \citet{Kasper2007,Kasper2012} to include more than 23 years.
This time period encompasses solar cycles 23 and 24 along with the end of solar cycle 22, thereby covering one Hale cycle.
In other words, an idealized sun with a pure dipole magnetic field would \added{have experienced two polarity reversals and} be returning to the configuration it had at the end of cycle 22.

In this work, we expand on the results of \citet{Kasper2007,Kasper2012}.
We show a positive correlation between \ahe\ and \ssn\ across multiple solar cycles.
In the slowest wind, we find a characteristic \ahe\ that is consistent across multiple minima and maxima.
Examining this relationship over one Hale cycle, we \replaced{find a clear time lag or phase delay between \ahe\ and \ssn}{demonstrate that the phase lag between \ahe\ and \ssn\ found by \citet{Feldman1978} is \replaced{an increasing}{a monotonically increasing} function of \vsw}.
This delay is characteristic to a given \vsw\ and, at any one \vsw, a cyclic delay is sufficient to correct for this lag.
\deleted{This lag is also an increasing function of \vsw.}
\added{Unexpectedly, \ahe\ returns to similar values in both maximum 23 and maximum 24 even though $\ssn_{\mathrm{Max},24} < \ssn_{\mathrm{Max},23}$.}
Our results are consistent when using the 13-month smoothed, monthly, and daily sunspot numbers.

The remainder of this Letter is dedicated to analyzing and interpreting this speed-dependent lag.
\sect{data} describes the observations and FC specifics that are key to this study.
\sect{abundance} describes the variation of \ahe\ with \vsw\ and \ssn\ over two solar cycles.
\sect{tlag} analyzes the delay in response of \ahe\ to changes in \ssn\  as a function of \vsw.
\sect{hysteresis} presents the relationship between \ahe\ and \ssn\ in various \vsw\ quantiles, corrected for the delay of peak cross-correlation coefficient.
Here, we show that correcting for the lag in \ahe's response to changes in \ssn\ reduces this hysteresis effect to a linear relationship.
In \sect{robust}, we use \ahe's dependence on \ssn\ to investigate the robustness of the \ahe, \vsw, \ssn\ relationship reported by \citet{Kasper2007}.
\added{In} \sect{hfilt}, we interpret our results and extend\deleted{s} earlier hypotheses regarding two sources of slow solar wind.
Finally, \sect{dnc} summarizes these results and discusses future work.

\section{Data Sources} \label{sec:data}

The \Wind\ spacecraft has been in continuous operation since its launch in the fall of 1994.
\citet{Ogilvie1995a} provide a detailed description of the Solar Wind Experiment (SWE) Faraday cups (FC).
\replaced{Multiple data products, optimized to resolve and measure different solar wind properties, have been produced.}{\citet{Kasper2006} introduce techniques for optimizing the algorithms that extract physical quantities from FC measurements. \citet{Maruca2013a} and \citet{Alterman2018a} build on these algorithms.}
\replaced{Throughout \Wind's mission,}These data have resulted high precision \added{solar wind} measurements of alpha particles \citep{Kasper2006,Maruca2013a} and multiple proton populations \citep{Alterman2018a}. \deleted{in the solar wind}
\replaced{In this paper, we use the data \replaced{of}{processed according to} \citet{Kasper2006}.}{The FC ion distributions are available on CDAweb\footnote{\url{https://cdaweb.sci.gsfc.nasa.gov/misc/NotesW.html\#WI_SW-ION-DIST_SWE-FARADAY}} and SPDF\footnote{\url{ftp://spdf.gsfc.nasa.gov/pub/data/wind/swe/swe_faraday/}}. We follow \citet{Alterman2018a} and reprocess the raw measurements to extract two proton populations (core and beam) along with an alpha particle population. The proton core is the population with the larger of the two proton densities. We calculate the solar wind speed as the proton center-of-mass velocity and treat the proton core as the proton density when calculating \ahe.}

Two aspects of FCs are key to this work.
First, FCs are energy-per-charge detectors.
In the highly supersonic solar wind, alpha particles and protons are well separated by the instrument even when they are co-moving \citep{Kasper2008,Kasper2017,Alterman2018a}, \added{as is commonly the case in slow solar wind}.
Second, the measurement quality has been stable and accurate throughout the mission \citep{Kasper2006}. \deleted{and \ahe\ is the ratio of two currents, the measurements used here are particularly robust}
These two FC characteristics enable our study of \ahe\ variation with a single dataset from one instrument suite covering the 23 years necessary to observe one Hale cycle.

\section{Solar Cycle Variation} \label{sec:abundance}

\cref{fig:abundance_cycle} presents \ahe\ as a function of \vsw\ and time over 23 years.
This time period starts at the trailing end of cycle 22 and extends through the declining phase of cycle 24.
\cref{fig:abundance_cycle} follows the format of \citet[Figure (1) in each]{Kasper2007,Kasper2012} and can be considered an update to their results.
The solar wind speed measurements from the full mission have been split into 12 quantiles.
The fastest and slowest quantile have been discarded due to measurement and statistical considerations.
Of those quantiles retained, the lower edge of the slowest is $312 \; \kms$ and the upper edge of the fastest is $574 \; \kms$.
Consequently, this study is limited to solar wind typically categorized as slow or slow and intermediate speed.\footnote{To be consistent with prior work (e.g.\ \citet{Kasper2007,Kasper2012}), we will use \emph{slow} and \emph{fast} to refer to the different extremes presented here. However, the reader should known that truly fast solar wind is excluded from our study.}
As in prior work, the abundance in each \vsw\ quantile is averaged into 250 day intervals.
The 13-month smoothed sunspot number \citep[\ssn]{sidc,Vanlommel2005} is interpolated to the measurement time; averaged into the same 250 day intervals as \ahe; and plotted on the secondary y-axis.
The legend indicates the middle of the solar wind speed quantile along with its corresponding Spearman rank cross correlation coefficient between \ahe\ and \ssn.
For brevity, we henceforth indicate the Spearman rank cross-correlation coefficient between \ahe\ and \ssn\ as \xcorr.

\cref{fig:abundance_cycle} indicates that \xcorr\ peaks at $\vsw = 355 \; \kms$.
The present drop in \ahe\ reflects that the sun is entering Minimum 25.
In contrast to the results of \citet{Kasper2007,Kasper2012}, $\xcorr > 0.6$ indicates a meaningful cross-correlation in all but the fastest reported quantile with $\vsw = 542 \; \kms$ and $\xcorr\ \geq 0.7$ is highly significant up to $\vsw = 426 \; \kms$. 
\added{As \citet{Feldman1978} noted,} there is \deleted{also} a phase offset between \ahe\ and \xcorr.
\added{Although the cycle 23 \ssn\ amplitude is less than the cycle 24 amplitude, \ahe\ unexpectedly returns to comparable values during each maximum.}

\section{Time-Lagged Cross Correlation} \label{sec:tlag}


Visual inspection indicates a clear time lag between \ahe\ and \ssn.
To quantify this lag, we calculate \xcorr\ as a function of delay time applied to \ssn\ from $-200 \; \days$ to $+600 \; \days$ in steps of 40 days--slightly longer than one solar rotation--for each \vsw\ quantile.
We smooth these results to reduce the impact of discretization.
The delay time is the time for which \xcorr\ peaks as a function of delay.
\added{Panel (a) of} \cref{fig:vsw_xcorr_delay_ebar} plots the peak cross correlation coefficient as a function of \vsw\ for observed (empty marker) and delayed (filled marker) \ssn.
Marker colors and symbols match \replaced{previous figure styles}{\cref{fig:abundance_cycle} and are maintained throughout the Letter}.
Dotted lines connect the markers to aid the eye.
To estimate the error in this calculation and its sensitivity to averaging timescale, we repeated it for averaging windows $\ndays = 225$ to $\ndays = 275$ in steps of 5 days.
Because a trend is not apparent, we choose to quantify this variability as the standard deviation across $\ndays$ and represent it as error bars centered on the $\ndays = 250$ averaging window utilized in this Letter.

Several features in \added{Panel (a) of} \cref{fig:vsw_xcorr_delay_ebar} stand out.
First, it emphasizes that delayed $\xcorr \geq 0.7$ is highly correlated for all \vsw\ quantiles.
Second, observed and delayed \xcorr\ peak at the same $\vsw = 355 \; \kms$.
Third, the change in \xcorr\ is largest and most visually striking in faster wind.
However, smaller changes in slower wind's \xcorr\ are statistically more significant because they are less likely to be due to random fluctuations.

\plotVswPeakDelay

\added{Panel (b) of} \cref{fig:vsw_xcorr_delay_ebar} examines $\tau$, the delay of peak \xcorr, as a function of \vsw.
\added{A positive delay indicates that changes in \ssn\ precede changes in \ahe.}
The insert at the top of the figure indicates the functional form, fit parameters, and quality metrics.
As with \added{Panel (a) of} \cref{fig:vsw_xcorr_delay_ebar}, the error bars indicate the variability of $\tau$ in each \vsw\ quantile.
Solving the fit equation for $\tau = 0$, or the speed at which \ahe\ responds immediately to changes in \ssn, results in $\vi = 200 \; \kms$.
Nevertheless, it is not unambiguously clear \deleted{unclear} if delay time $\tau$ monotonically increases with \vsw\ or there are two distinct delay times.
If it is actually the latter, then \ahe\ in slow wind responds to changes in \ssn\ with a delay time $\tauS = 150 \; \days$; faster wind responds after $\tauF > 300 \; \days$; and  \vi\ represents a non-trivial conflation of these two delays.
If this is not the case, it may be that \tauS\ is the shortest delay with which \ahe\ responds to changes in \ssn. \added{As discussed below, in either case} all helium released into the solar wind still lags changes in \ssn.

\section{Phase Delay} \label{sec:hysteresis}

\cref{fig:hysteresis} presents \ahe\ as a function of \replaced{sunspot number \ssn}{\ssn\ in} the example quantile \replaced{$\vsw =  426 \; \kms$}{$\vsw =  355 \; \kms$}.
\replaced{Although smaller changes in higher \xcorr\ are more significant, we show a \vsw\ quantile for which the change in cross-correlation coefficient $\Delta \xcorr$ is sufficiently large to illustrate the phase delay's effect.}{This is the \vsw\ quantile for which the change in cross-correlation coefficient $\Delta \xcorr$ is smallest and the phase delay's effect is least likely to be due to random fluctuations.}
Panel (a) uses the observed \ssn. 
Panel (b) uses \ssn\ delayed by the time indicated in \added{Panel (b) of} \cref{fig:vsw_xcorr_delay_ebar}, \replaced{$\sim 200 \; \days$}{$\sim 150 \; \days$}.
A line connects the points to aid the eye.
Both line and marker color indicate the days since mission start, given by the color bars.
\added{Marker shapes match the style of previous figures.}
Both panels contain a robust fit to the data, each indicating the monotonic, increasing trend. 
As in \added{Panel (b) of} \cref{fig:vsw_xcorr_delay_ebar}, the insert at the top of each panel describes the fit.

Panel (a) clearly shows the hysteresis pattern of \ahe\ as a function of \ssn.
As seen with other indices (e.g.\ \citet{Bachmann1994}), \deleted{each cycle's rising edge is below the trend line and its falling edge is above it. In other words,} time moves counter-clockwise in this \replaced{panel}{plot}.\footnote{Not all indices present with the same handedness and the handedness of some changes across solar cycles \citep{Ozguc2001}. A larger study of \ahe\ variation is necessary to generalize this handedness observation.}
As noted by \citet{Bachmann1994} for several solar indices, the clustering of data at small \ssn\ indicates that the hysteresis effect is stronger at solar maximum and weaker at solar minimum.

In panel (b), the larger \rsq\ indicates that this spread of \ahe\ about the trend decreases.
Note that \rsq\ corresponds to the square of the correlation coefficient of  \ahe\ and \ssn\ derived from a robust fit and not directly from the measurements.
\emph{Although $R$ is similar to \xcorr, they are not trivially equal.}
That delayed \chisqnu\ is markedly closer to unity indicates that a linear model better characterizes \ahe\ as a function of delayed \added{rather} than observed \ssn.
Because delayed \ssn\ only reduces the spread of \ahe\ about the trend, it is expected that the trends \added{and fit parameters} in both cases are similar.

\plotHysteresis

\section{Robustness of $\ahe(v)$} \label{sec:robust}

\citet{Kasper2007} describe the relationship between \ahe\ and \vsw\ in slow wind ($\vsw \leq 530 \; \kms$) using data from a 2 year interval surrounding solar Minimum 23.
They find that $\ahe(v) = 1.63 \times 10^{-2} \, (v - v_0)$, where $v_0 =  259 \pm 12 \; \kms$ is the speed below which helium vanishes from the solar wind.
The robust fits in \cref{fig:hysteresis} allow us to extract \ahe\ at zero solar activity for all \vsw\ quantiles. This quantity, $\ahe(\ssn = 0)$, represents low solar activity conditions across this Hale cycle that are appropriate for comparison to the minimum 23 results from \citet{Kasper2007}.

\cref{fig:hysteresis-summary} plots $\ahe(\ssn = 0)$ in all \vsw\ quantiles for delayed \ssn\ with unfilled markers.
As observed \ssn\ does not deviate from delayed \ssn\ in this figure, it is omitted for clarity.
The black dashed curve is the fit of $\ahe(v)$ from \citet{Kasper2007}. 
To better compare this analysis to the work of \citet{Kasper2007}, filled markers present the results of repeating this analysis for $\ssn < 25$, a range in \ssn\ representative of solar minimum 23.
That \replaced{subset of data better aligns with their results--especially in slow wind--}{$\ahe(\ssn = 0)$ is smaller in this reanalysis using a restricted range of \ssn}  further substantiates that our results are consistent with those of \citet{Kasper2007} even though ours cover multiple solar cycles\replaced{ and}{, } a larger range in solar activity conditions, \added{ and uses a different analysis technique}.
\added{Furthermore, the agreement between these two distinct analysis techniques supports the interpretation that helium release is essential to solar wind formation \citep{Kasper2007}.}
The discrepancy between our fastest quantile with $\vsw = 542 \; \kms$ and their trend is expected because (1) it is outside of the speed range they fit and (2) they found thet \ahe\ at this and similarly high speeds takes on a stable value between $4\%$ and $5\%$.

\plotHysteresisSummary

\section{Helium Filtration}\label{sec:hfilt}


Many solar indices have a distinct phase-offset or hysteresis-like behavior with \ssn\ \citep[and references therein]{Ramesh2014}.
Two such indicators include Lyman-$\alpha$ (\Lalpha) intensity and soft x-ray flux (\sxr).
\Lalpha\ measures activity in the sun's chromosphere \& transition region \citep{Fontenla2001,Fontenla1988} and lags \ssn\ by 125 days \citep{Bachmann1994}.
\sxr\ \deleted{is generated by solar flares \citep{Benz2008,Temmer2003},} is most common in Active Regions (AR) \citep{VanDriel-Gesztelyi2015}, and lags \ssn\ by 300 days to 450 days \citep{Temmer2003}.


While \ahe\ is approximately $8.5\%$ within the sun's convection zone and out to the photosphere \citep{Asplund2009,Laming2015a}, it rarely exceeds $5\%$ in the corona \citep{Laming2003,Mauas2005}. 
It has long been assumed that \ahe\ is initially modified in the photosphere.
However, the speed-dependent lag in \ahe's response to changes in \ssn\ found here suggests additional processes at higher altitudes further modify helium's abundance.
Slow solar wind's 150 day lag tracks lags in transition region and chromosphere structures, while faster wind's 300 day lag is more consistent with higher altitude structures in the corona.
How could the transition region or corona modify the helium abundance?

\citet{Kasper2007} propose that two mechanisms release fully ionized helium into the slow solar wind\added{, one each in the streamer belt and ARs}.
ARs have a strong magnetic field that extends from the photosphere into the corona, originate well above the equatorial region, tend to migrate towards the equator as they get older, and have loops that tend to grow with age \citep{VanDriel-Gesztelyi2015}.
In contrast, the streamer belt has a weaker magnetic field, is composed of loops larger than those typical of ARs, is magnetically closed to the heliosphere, and is typically considered the source of slow solar wind \citep{Eselevich2006}.
\citet{Stakhiv2016} identify signatures of these two solar sources in ACE/\SWICS\ composition measurements.

If there are two sources of slow wind, solar wind originating in the streamer belt is more processed than that originating in ARs, where \sxr\ is enhanced.
Slower wind \ahe\ ($\vsw < 375 \; \kms$) originates from the streamer belt with a phase delay $\tauS = 150 \; \days$.
It appears more depleted than faster solar wind from ARs that has a phase delay $\tauF > 300 \; \days$.
The magnitude of \ahe's reduction from its photospheric value and the speed-dependent delay then reflect the extent to which a given source region is magnetically open to the heliosphere.
As the phase delay between \ahe\ and \ssn\ is an increasing function of \vsw, ARs and the streamer belt may be two extreme cases along the continuum of slow wind helium depletion mechanisms.



For \replaced{illustration}{illustrative} purposes, one candidate mechanism that may contribute to this processing is the FIP effect.
The FIP effect is the empirical observation that solar wind ion\added{s} \deleted{abundances} \replaced{are either enhanced or depleted with respect to}{are fractionated, or their abundances differ from} their photospheric value based on their first ionization potential \citep[and references therein]{Meyer1991,Meyer1993,Laming2015a}.
\added{Low FIP elements ($\mathrm{FIP} < 10 \; \mathrm{eV}$) tend to increase or experience an enhancement. This low-FIP enhancement also leads to an apparent depletion in high-FIP elements, as with helium.}
\deleted{It can be explained as the result of interactions between the pondermotive force driven by coronal Alfv\'en waves and magnetic loops \citep{Rakowski2012}.}
\added{Under the framework of \citet{Rakowski2012}, time averaged coronal Alfv\'en waves create a pondermotive force that accelerates ions into the corona and leads to fractionation in coronal loops.}
\replaced{It}{The FIP effect} is strongest in the upper chromosphere and lower transition region, weakest in regions of strong magnetic field, \added{and} stronger in longer loops\deleted{, and leads to a depletion of helium } \citep{Rakowski2012}.
\added{\citet{Feldman2005a} found that FIP bias in ARs increases with age.}

However, this is just one of several possible mechanisms that could cause this phase lag. 
Other mechanisms that might impact the speed-dependent phase lag may include interchange reconnection \citep{Fisk2003} and gravitational settling \citep{Hirshberg1973,Borrini1981,Vauclair1991}.
\added{Moreover, these are mechanisms are not mutually exclusive. \citet{Schwadron1999,Laming2004,Rakowski2012} include gravitational settling in their models of the FIP effect.  \citet{Schwadron1999} also relies on interchange reconnection to create the magnetic structures necessary for FIP fractionation to occur. As \citet{Rakowski2012} show, the combination of coronal loop length, differences in gravitational scale height, and the FIP effect can lead to the apparent depletion of \ahe.}
Whatever the underlying mechanism, it should also account for the observation that \ahe\ returns to a similar value during solar maximum, irrespective of \ssn\ during maximum.


\section{Conclusion} \label{sec:dnc}

Following the methods of \citet{Kasper2007,Kasper2012}, we have analyzed the relationship between \ahe\ and the 13-month smoothed sunspot number (\ssn) by studying their cross correlation coefficient using 250 day averages.
We have verified that our results are consistent when using the monthly and daily \ssn.
Our data covers 23 years, including cycle 23 and 24 along with the tail end of cycle 22.
This time period is more than the 22 years of a Hale cycle over which the pure dipole field of an idealized sun would \added{experience two polarity reversals and} return to an initial configuration.
\replaced{The}{As shown in \cref{fig:abundance_cycle}, the} present decrease in \ahe\ clearly demonstrates that we are entering solar Minimum 25.
While the significance of \added{the cross correlation coefficient} \xcorr\ decreases with increasing \vsw, \cref{fig:abundance_cycle} shows that \xcorr\ is meaningful up to $\vsw = 488 \; \kms$ and highly significant up to $\vsw = 426 \; \kms$.
\added{A subject of future work is investigating why \ahe\ returns to a similar value in Maximum 24 even though cycle 24's amplitude is markedly smaller than cycle 23's.}

\replaced{These 23 years of measurements revealed a phase offset between \ahe\ and \ssn, which appears as a hysteresis.}{\citet{Feldman1978} comment on a phase offset between \ahe\ and \ssn.}
\replaced{Three key results flow from Panel (a) of Fig.\ (\ref{fig:vsw_xcorr_delay_ebar}). (1) Changes in \ssn\ precede changes in \ahe. (2) The length of this delay is an increasing function of \vsw. (3) The \vsw\ quantile most correlated with \ssn\ does not change when \ssn\ is appropriately delayed in each \vsw\ quantile.}{Panel (b) of \cref{fig:vsw_xcorr_delay_ebar} reveals that (1) the length of this delay is an increasing function of \vsw\ and (2) the \vsw\ quantile most correlated with \ssn\ does not change when \ssn\ is appropriately delayed in each quantile.}
We have also argued that, although changes in \xcorr\ are most dramatic in faster \vsw\ quantiles, the probability of smaller changes in slower wind's larger \xcorr\ is much smaller and therefore more significant.

\added{Panel (b) of} \cref{fig:vsw_xcorr_delay_ebar} presents the delay applied to \ssn\ necessary to maximize \xcorr\ as a function of \vsw.
\added{The delay is a monotonically increasing function of \vsw\ and} linear fit to this trend \replaced{and found}{reveals} that the speed at which \ahe\ responds instantaneously to changes in \ssn\ is $\vi = 200 \; \kms$.
Yet the speed of instantaneous response is less than the vanishing speed, $\vi < \vv$.
Therefore any helium released into the solar wind will necessarily response to changes in \ssn\ after some delay. 
If trend in \added{Panel (b) of} \cref{fig:vsw_xcorr_delay_ebar} is correct, then the minimum delay in \ahe's response to \ssn\ is $68 \pm 13 \; \days$, or approximately two Carrington Rotations.
Here, we also note that there may be two distinct phase delays (\tauS\ and \tauF) with which \ahe\ responds to changes in \ssn\ and the fit quantity \vi\ may be a conflation of the physics related to each phase delay.
Under either interpretation, helium released into the solar wind is a delayed response to changes in \ssn.

In \sect{hysteresis}, we present robust fits to \ahe\ as a function of observed and delayed \ssn\ in each \vsw\ quantile.
It visually illustrates that applying a time delay to \ssn\ reduces the spread of \ahe\ about its trend.
In \sect{robust}, we use helium abundance at zero solar activity derived from these \deleted{robust} fits to demonstrate that our results using 23 years of data are consistent with the trend found by \citet{Kasper2007} for a two year interval surrounding solar minimum 23.

In \sect{hfilt}, we discuss how the demonstrated phase delay or hysteresis effect is qualitatively similar to the phase delays between \ssn\ and many regularly observed solar indices \citep[and references therein]{Ramesh2014}.
We \replaced{argue}{note} that the two aforementioned phase delays \added{(\tauS\ and \tauF)} are consistent with \Lalpha\ and \sxr\ and that this consistency is indicative of two distinct source regions.
Slower wind ($\vsw < 375 \; \kms$) with a lower \ahe\ originates in the streamer belt \deleted{, is depleted or filtered by the FIP effect,} and responds to changes in \ssn\ with characteristic delay time $\tauS = 150 \; \days$.
Faster wind with a larger \ahe\ originates in ARs \deleted{, is impulsively released by the same flare-related mechanism that causes \sxr, is not as FIP depleted,} and responds to changes in \ssn\ with characteristic delay time $\tauF > 300 \; \days$.
\added{These different delay times indicate that \ahe\ is processed by one or more mechanisms above the photosphere.}
\replaced{Assuming that \citet{Kasper2007} are correct}{Assuming that the results of \citet{Kasper2007} apply across the solar cycle} and helium \added{universally} vanishes from the solar wind when $\vsw < 259 \; \kms$ \added{irrespective of solar activity}, one possible interpretation is that \added{there is a minimum \ahe\ necessary for solar wind formation,} the mechanisms that \replaced{filter}{reduces} \ahe\ to a value less than its photospheric value \replaced{fully filter \ahe}{prevents solar wind release} below the vanishing speed \replaced{\vi}{\vv}, and\replaced{,}{--}using the fit from Panel (b) of \cref{fig:vsw_xcorr_delay_ebar}\replaced{,}{--}any helium that enters the solar wind is released after 68 days, \deleted{or} approximately two Carrington rotations.
\added{If this is the case, helium in the high-speed solar wind may represent the solar wind's ``ground state'' \citep{Bame1977,Schwenn2006} and  the observed depletion of \ahe\ is the result of source regions departing from states that release fast wind, i.~e.~those magnetically open to the heliosphere.}
A rigorous study of the relationship between \ahe\ and solar indices other than \ssn\ may better constrain helium variation by source region \added{and} is a subject of future work.

This work highlights the value of recent and forthcoming advances in heliophysics.
Parker Solar Probe \citep[PSP]{Fox2015} launched in August, 2018 and completed its first perihelion in November of that year. 
Solar Orbiter \citep[SolO]{Muller2013} will launch in 2020.
The \added{thermal ion} instruments on board \replaced{(e.g.\ \citet{Kasper2015})}{\citep{Kasper2015}} provide an unprecedented opportunity to study the solar wind, its formation, and its acceleration.
For example, PSP will make measurements near and below the Alfv\'en critical point, i.e.\ at distances within which mapping the solar wind to specific sources is significantly simplified in comparison with \Wind.
\citet{McMullin2016} anticipate that the Daniel K. Inouye Solar Telescope (DKIST) will begin operations in 2020.
DKIST's Cryo-NIRSP instrument will be capable of simultaneously imaging solar helium at various heights in the corona.
Combining DKIST measurements with PSP and SolO measurements will enhance our ability to differentiate between the mechanisms releasing helium into the solar wind--e.g.\ from the streamer belt or ARs--and better constrain the delay in helium's response to changes in \ssn.

\acknowledgments{This work was funded by NASA grants NNX17AI18G and 80NSSC18K0986. We are grateful to Michael Stevens for supplying the data and the referee for insightful comments. We also thank Lennard Fisk, Enrico Landi, Liang Zhao, Phyllis Whittlesey, and Michael Stevens  for fruitful discussions.}

\software{
IPython \citep{Perez2007}, 
Jupyter \citep{Kluyver2016a}, 
Matplotlib \citep{Hunter2007}, 
Numpy \citep{VanderWalt2011}, 
Pandas \citep{Mckinney2010}, 
Python \citep{Oliphant2007,Millman2011}
}

\bibliography{Mendeley.bib}   


\end{document}